\documentclass[aps, prb, twocolumn, showkeys, floatfix]{revtex4-1}

\usepackage{chemformula} 
\usepackage{siunitx} 
\usepackage[inline,shortlabels]{enumitem} 
\usepackage{hyperref} 
\usepackage{graphicx} 
\usepackage{tabularx}
\usepackage{threeparttablex} 
\usepackage[section]{placeins}
\usepackage{multirow}
\usepackage{fancyhdr}
\usepackage{dblfloatfix}
\usepackage{float}

\hypersetup{hidelinks}

\makeatletter
\renewcommand{\fnum@figure}{Fig. \thefigure}
\makeatother

\usepackage[justification=raggedright, labelsep=period]{caption}

\begin{document}

    \title{Polymers’ surface interactions with molten iron: a theoretical study}
    \author{M. Hussein N. Assadi}
    \affiliation{Centre for Sustainable Materials Research and Technology (SMaRT), School of Materials Science and Engineering, The University of New South Wales, Sydney, 2052, Australia}
    \affiliation{Graduate School of Engineering Science, Osaka University, Toyonaka, Osaka 560-8531, Japan.\\
    \url{assadi@aquarius.mp.es.osaka-u.ac.jp}}
    \author{Veena Sahajwalla}
    \affiliation{Centre for Sustainable Materials Research and Technology (SMaRT), School of Materials Science and Engineering, The University of New South Wales, Sydney, 2052, Australia.}
    \date{2014}
    
    \begin{abstract}
    Environmental concerns are the chief drive for more innovative recycling techniques for end-of-life polymeric products. One attractive option is taking advantage of C and H content of polymeric waste in steelmaking industry. In this work, we examined the interaction of two high production polymers, \textit{i.e.}, polyurethane and polysulfide with molten iron using \textit{ab initio} molecular dynamics simulation. We demonstrate that both polymers can be used as carburizers for molten iron. Additionally, we found that light weight \ch{H2} and \ch{CH_x} molecules were released as by-products of the polymer-molten iron interaction. The outcomes of this study will have applications in the carburization of molten iron during ladle metallurgy and waste plastic injection in electric arc furnace.
    \end{abstract}

\maketitle

       \section{INTRODUCTION}
The annual global production of polymeric goods is expected to exceed 300 million tons in 2014 [1]. With only $\sim 50 \%$ of that amount being currently recovered [2], the generated waste by end-of-life polymeric goods exerts paramount pressure on our ecosystem. The generation of micro-plastic oceanic soup and methane emission from degrading plastics in landfills are among the main challenges [3]. As a result, the environmental concerns have prompted new regulations in various jurisdictions aiming at reducing plastic waste in the landfills [4--6]. One innovative approach is to utilize the vast amount of H and C embedded in polymers as energy resources and reductants in steelmaking industry. Japanese steelmakers have pioneered a recycling technology for blast furnaces that has been in operation since early 2000s [7]. More recently, our group has successfully proven the feasibility of using different types of waste polymers in electric arc furnace (EAF) steelmaking, particularly in slag foaming and carburization of steel during ladle metallurgy [8]. \par
Despite the experimental advances, there are many interesting theoretical questions regarding the interactions of polymers and molten Fe that are yet to be answered. For instance, what are the atomistic mechanisms that govern the dissolution of certain elements in the molten Fe and the emission of others as volatiles? Or more importantly, how the presence of the other elements affects the dissolution rate of C in molten Fe? The answer to this question establishes the suitability of waste polymeric products as carburizing agents. We know that dissolution kinetics is not only determined by the diffusivity of a given element into the molten Fe but also by the presence of other elements too. This is due to the possible formation of a barrier layer at the liquid-solid interface that can hinder the dissolution process. The formation of a barrier layer is expected to have more complicated effects in the case of carburizing molten Fe using polymers as polymers come in variety of chemical compositions containing elements such as H, O, N and S in addition to C. These elements each have different diffusion rates in molten Fe, thus accumulate in different concentrations on the interface and subsequent layers. In laboratory settings, rotating disc experiments have been widely used to study the dissolution of various materials in molten metals. This method, however, is not applicable to the case of polymers as they have very low melting temperatures, thus they combust or gasify rather quickly. Consequently, we should resort to theoretical methods to obtain fundamental understanding of the polymer-molten Fe interaction and thus establish their suitability for steelmaking processes. Particularly \textit{ab initio} molecular dynamics (AI-MD) offers a very flexible and reliable approach in obtaining a unified picture of diffusion process and volatiles formation. \par
In this work we examine two polymers interacting with molten Fe using AI-MD simulation. The first polymer was the thermoset polyurethane which is widely used in consumer products. Polyurethane contains a combination of carbon, hydrogen, oxygen and nitrogen in a range of concentrations such as \ch{C25H35NO3}, \ch{C15H21NO5} etc. Currently there is no economical methods to recycle polyurethane in plastic form other than costly mechanical or chemical recycling techniques [9]. Therefore, significant quantities of end-of-life polyurethane end up in landfills. The second polymer was polysulfide which is used as an elastomer in synthetic rubber specifically used in passenger vehicle tires. Polysulfide interconnects neighboring carbon chains and thereby conferring rigidity. As a result, a typical passenger vehicle tire contains up to 1 wt.$\%$ S due to the polysulfide additives [10]. Characterizing the interactions of polyurethane and polysulfide with molten Fe, will open a new avenue for reducing the adverse environmental impact when these polymers reach the end-of-life cycle.

       \section{METHODOLOGY}
       \subsection{Computational Settings}
Spin polarized \textit{ab initio} molecular dynamics (AI-MD) were performed with the Vienna \textit{ab initio} simulation (VASP) package [11--13] using plane-wave basis for expanding electronic wave functions. Electronic exchange and correlation interactions were approximated by PW91 functional [14]. The energy cutoff for all configurations was set 500 eV while only gamma point was used to generate the k-point grid. The self-consistency tolerance threshold was set at $10^{-5}$ eV. AI-MD simulations were performed using Nose--Hoover thermostat to produce an NVT ensemble [15, 16]. The time step for all molecular dynamics was set, at most, to 0.5 fs. The simulation was performed on Intel Xeon Sandy Bridge 2.6 GHz processors and consumed $\sim 2 \times 10^{+5}$ hours of CPU time. \par
       \subsection{System Settings}
\begin{figure}[bth!]
           \centering
            \includegraphics[width=1\linewidth]{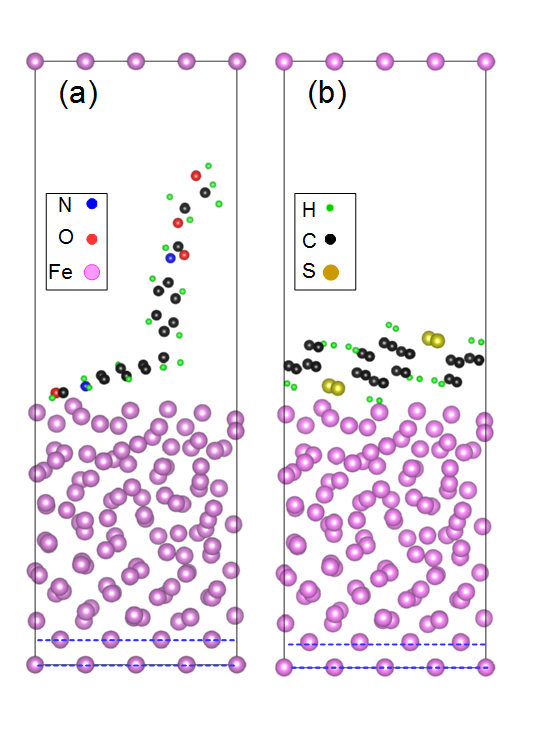}
            \caption{\label{figure:1}The initial structures of the (a) polyurethane and (b) polysulfide monomers in interaction with molten Fe. The dashed blue lines indicate the Fe planes that were fixed during the AI-MD simulations.}
         \end{figure}

Initially a supercell of $3a \times 4b \times 5c$ of conventional $\alpha$Fe cell containing 132 Fe atoms was constructed. The dimension of this supercell was 8.59 $\mathrm{\AA}$ $\times$ 11.46 $\mathrm{\AA}$ $\times$ 14.32 $\mathrm{\AA}$. This structure was then cleaved along (0 0 1) plane and a vacuum slab of 20 $\mathrm{\AA}$ was added. The large vacuum space was required to study the volatile that might form during the interactions. Then, the first two planes of Fe with $z = 0$ $\mathrm{\AA}$ and $z = 0.9$ $\mathrm{\AA}$ were fixed while the rest of Fe atoms were allowed to equilibrate at 1823 K for 20 ps to reach the liquid phase. The fixed Fe layers (marked in Fig. \ref{figure:1}) were required to prevent artificial interactions and diffusion of volatiles into molten Fe at the second Fe interface. We have previously demonstrated that the obtained molten Fe's pair correlation function was in good agreement with experimental results [17]. \par

\begin{figure}[bth!]
            \centering
            \includegraphics[width=1\linewidth]{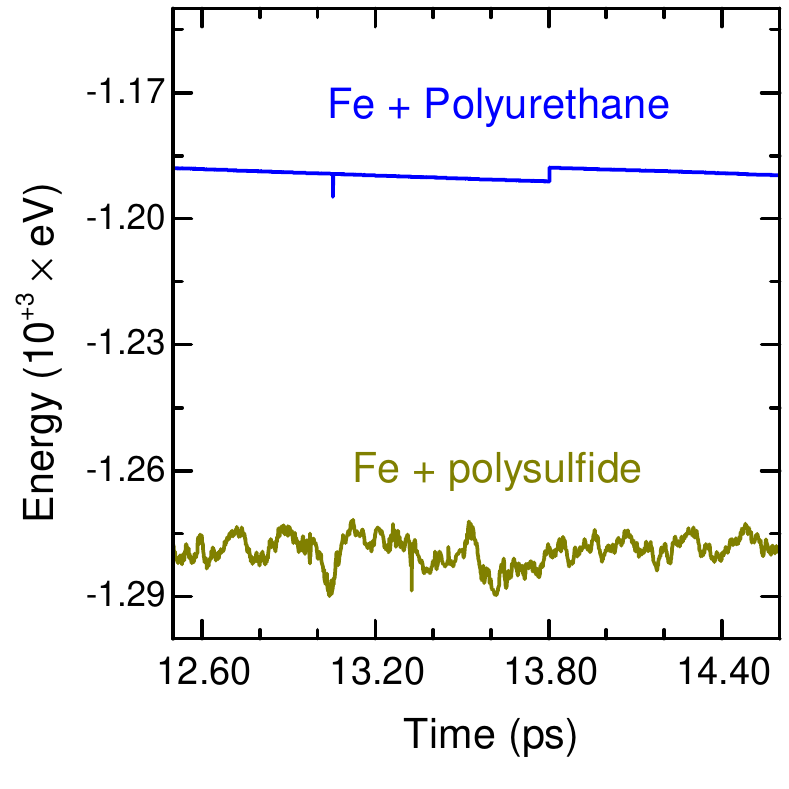}
            \caption{\label{figure:2}The energy fluctuations of molecular dynamics run in the final stages of the simulation. The larger magnitude of the fluctuations in the Fe plus polysulfide system is due to larger mass of S and \ch{CH_x} volatiles.}
         \end{figure}

 To study the dissolution of polyurethane and polysulfide in molten Fe, one single monomers of the respective polymers was placed at the interface of the molten Fe. The composition of the monomers is given in Table \ref{table:1}. A schematic representation of the initial atomic positions of the interacting monomers and molten Fe is also demonstrated in Fig. 1(a) and (b). In the both cases, few different initial configurations were examined to investigate the effect of the initial coordination on the simulation process. It was found that an approximate distance of $\sim 1.5$ $\mathrm{\AA}$ between the surface of molten Fe and a parallel and the monomers, as demonstrated in Fig. \ref{figure:1}(a) and Fig. \ref{figure:1}(b) would best facilitate the interactions. Then the monomer-molten Fe systems where allowed to equilibrate for 15 ps (30000 single point calculations for each monomer system) to ensure that energy fluctuations were smaller than $\sim 0.2 \%$ of total energy as demonstrated in Fig. \ref{figure:2}.
 
           \begin{table}[bth!]
\caption{The initial composition of the monomers and the interaction resultants.}
\begin{center}
\begin{threeparttable}
\begin{tabular}{m{3.7cm} m{2.7cm} m{1.8cm}}
  \hline\hline
           &  Polyurethane  & Polysulfide \\
  \hline
      Monomer Composition   &   \ch{C17H18O4N2}\tnote{1} & \ch{C24S4H16} \\
      No. of diffused C atoms &  3 & 10 \\
      No. of diffused H atoms & 3 & 4 \\
      Interface composition\tnote{2} & 1O--1N--3H--4C & 2S--3H--5C \\
      Volatiles & \ch{H2} & \ch{H2}, \ch{CH_x}, S \\
  \hline\hline
\end{tabular}
\begin{tablenotes}
\item[1] Both ends of the polyurethane monomer were saturated with hydrogen to maintain charge neutrality. \item[2] The presented composition does not include the Fe atoms that are readily present in the interface region.
\end{tablenotes}
\end{threeparttable}
\end{center}
\label{table:1}
\end{table}
 
       \subsection{Analysis of Data}
The interaction of monomers with molten Fe was analyzed by probing the density profiles ($\rho(z)$ ) for each element sampled over series of bins that divided the supercell. These bins are set to be parallel to the molten Fe surface \textit{e.g.} $z$ direction.  $\rho(z)$ for a given element was calculated using the following equation:
       \begin{equation} \label{eq1}
\rho (z) = \frac{\langle N_z \rangle}{A_{xy}\Delta_z}
       \end{equation}
Here $\langle N_z \rangle$ is the time-averaged number of atoms in a given bin of the length of  and $A_{xy}$ is the bins' cross section area. In order to obtain accurate spatial density profile, the bin size $\Delta_z$ were chosen in a manner that allows the density variation within the bin to be far smaller than the density variation across neighboring bins. Therefore, we divided the supercell along z direction into 343 bins for which  was equal to $\sim 0.1$ $\mathrm{\AA}$. The time average was taken over 100 snapshots separated by 0.05 fs. The time interval between these snapshots was set an order of magnitude smaller than the one of AI-MD simulation time steps to avoid averaging over the general evolution of monomer-molten Fe interaction. The abundance of different elements then was probed in three distinct regions. First, the bulk area of molten Fe where the presence of an element indicated diffusion. Second, the interface layer which consists of the top outmost Fe atoms where the contact between monomers and molten Fe occurred. Finally, volume above the interface in which volatile by-products were confined.
             
       \section{RESULTS AND DISCUSSION}
       \subsection{Interfacial Composition and Volatiles}
\begin{figure}[bth!]
            \centering
            \includegraphics[width=1\linewidth]{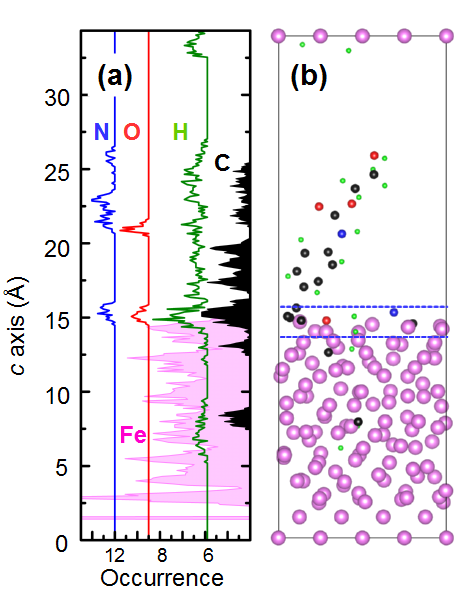}
            \caption{\label{figure:3}(a) Calculated density profiles of the diffusion process of the polyurethane monomer in molten Fe at $T = 1823$ K after 15 ps. (b) The schematic representation of final structure with averaged coordinates over 100 steps. The presentation of atoms are as following: (i) pink spheres for Fe, (ii) black spheres for C, (iii) green spheres for H, (iv) blue spheres for nitrogen and red for oxygen. The dashed lines mark the interfacial layer.}
         \end{figure}

 The density profile of the equilibrated structure of polyurethane monomer and molten Fe is presented in Fig. \ref{figure:3}(a) which indicates the C and H diffusion into molten Fe. More quantitatively as indicated in Table \ref{table:1}, out of the 17 C atoms three were diffused into molten Fe while four C atoms were deposited on the interface. Moreover, out of 18 H atoms, three were diffused in molten Fe while another three H atoms were deposited on the interface. As illustrated in Fig. \ref{figure:3}(b), by $t = 15$ ps, only one of the carbon rings of polyurethane's monomer was consumed by molten Fe. Consequently, the residual compound consisted of one dehydrogenated carbon ring attached to a C-N chain. The only volatile emission was gaseous H as indicated by the H peak at $z = \sim 34$ $\mathrm{\AA}$ indicates in Fig. \ref{figure:3}(a). In addition to C and H, one N atom and one O atom were also found to be deposited on the interfacial surface. However, no N atom nor any O atom was diffused into the bulk region of the molten Fe. The lack of N diffusion can be attributed to retardation of N dissolution in presence of C that has higher diffusivity in molten Fe. The retardation of N dissolution in molten Fe is in the presence of other elements has been previously reported in various experiments [18]. Furthermore, the preference of N atom to remain on the interfacial surface infers N aggregation at the grain boundaries in solidified steel. Such aggregation has also been observed experimentally on numerous occasions. Interfacial N atom can enhance the mechanical properties of various type of steels [19]. The reason is attributed to N's increasing the inter-planar friction among grain boundaries thus consequently hardening the steel. The interfacial O atom on the other hand is expected to be captured by the slag that would form during polymer-molten Fe interaction. \par
The equilibrated structure of polysulfide monomer and molten along with the corresponding density profile are Fe is demonstrated in Fig. \ref{figure:4}(a) and (b) respectively. Fig. \ref{figure:4} indicates that out of 24 C atoms, ten C atoms were diffused into the molten Fe while five C atoms were deposited on the interface. Additionally, out of 16 hydrogen atom, four hydrogen atoms were diffused in molten Fe while another three H atoms were deposited on the interface. In addition to the interfacial H and C, there were two S atoms on the interfacial region as well. The deposition of S at the interface was previously predicted using Monte-Carlo simulation [20] and was found to slow down the rate of carburization and decarburization of molten Fe. The emitted volatiles were gaseous \ch{H2}, S and nine \ch{CH_x} unsaturated hydrocarbons as indicated by the H, S and C peaks at  $z = \sim 34$ $\mathrm{\AA}$ in Fig. \ref{figure:4}(a). In this case, the emitted \ch{H2} and \ch{CH_x} can be used as foaming and reducing agents while S is expected to be absorbed by the slag. 

       \subsection{Diffusion Process}
Among all elements present in both monomers, H and C were the only elements that crossed the interface region and diffused into the bulk region of the molten Fe within the time scale of the simulation. We now try to analyze the diffusion process of H and C in detail and relate them to the macroscopic phenomena. Earlier experimental reports suggest that the dissolution of C consists of two steps: C's interfacial diffusion and then mass transport by turbulent liquid Fe motion [21]. The interplay between these two steps restricts the excessive C dissolution in molten Fe. Contrarily, H's dissolution in molten Fe is predominantly controlled by diffusion at the interface only, therefore significant quantities of H can be dissolved in stagnant molten Fe, even in the absence of turbulent motion of the molten Fe [22]. Through AI-MD simulation, we can obtain quantitative insight into the initial diffusion of the H and C dissolution in molten Fe while mass transport phenomenon related to turbulent liquids are out of the scope of our calculations.       
       
       \subsection{Carbon Diffusion}
To compare the carburization performance of polyurethane and polysulfide, we now calculate the \emph{diffusivity} ($D$) of C from the monomers using the Noyes-Whitney equation: $d \mathrm{C} / dt = \nicefrac{A D / d}{\rho_s - \rho_b}$. Here $A$ is the cross section of the simulation supercell, $d$ is the width of the interfacial region, $\rho_s$ is the concentration of C on the surface layer and $\rho_b$  is the C concentration in molten Fe. By having the volumes of the interface layer and the inner volume of the molten Fe, we then can calculate $\rho_s$ and $\rho_b$ from the C concentration in the interfacial and bulk areas. We can assume that  $d \mathrm{C} / dt = \mathrm{C}_b / \Delta t$ where C$_b$ in the number of C atoms in the bulk region and $\Delta t = 15$ ps. We determined if an atom was in the interface or bulk region by examining its coordination number based on bonding with Fe atoms. Bond connectivity was established by examining the charge density profiles of the averaged coordinates that were previously used to generate the density profiles. This procedure yields a C diffusivity of 0.105 $\mathrm{\AA}^2$ ns$^{-1}$ for polyurethane and 0.137 $\mathrm{\AA}^2$ ns$^{-1}$ for polysulfide. The calculated diffusivities indicate that by the time the monomer-molten Fe system had reached equilibrium, the mass ratio (weight percentage) of dissolved C was $0.49 \%$ for polyurethane and $1.68 \%$ for polysulfide. In the polysulfide case the dissolved C mass ratio was larger than that of polycarbonate, \textit{i.e.}, $1.45 \%$ which is one of the most efficient carburizing polymers for molten Fe [17]. The lower carburization performance of polyurethane may be attributed to its lower density and also the presence of N and O in its composition.       

           \begin{figure}[bth!]
            \centering
            \includegraphics[width=1\linewidth]{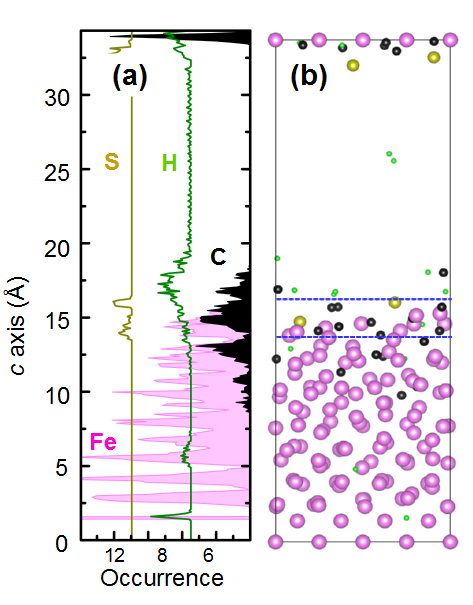}
            \caption{\label{figure:4}Calculated density profiles of the diffusion process of the polysulfide monomer in molten Fe at $T = 1823$ K after 15 ps. The isolated H and S peaks in (d) at $z = \sim 34$ $\mathrm{\AA}$ denote isolated gaseous H and S that are detached from the polysulfide monomer. (b) The schematic representation of final structure with averaged coordinates over last 100 steps. The presentation of atoms are as following: (i) pink spheres for Fe, (ii) black spheres for C, (iii) green spheres for H, (iv) dark yellow spheres for sulfur. The dashed lines mark the interfacial layer.}
         \end{figure}
             
       \subsection{Hydrogen Diffusion}
 For both monomers we found that hydrogen was present in three forms: (i) hydrogen diffused through the interface to the bulk region; (ii) the interfacial hydrogen and (iii) the volatile hydrogen. The continuous fluctuations of H density profiles in Fig. \ref{figure:3}(a) and Fig. \ref{figure:4}(a) over these three regions indicate the highly mobile nature of H at this temperature range. The volatility of H at $T = 1550$ $^{\circ}$C, and its effects on metallurgical processes has been widely studied by experimentalists [23]. The solubility of H in molten Fe is, therefore, known to be critically dependant to \ch{H2}'s partial pressure [24] and consequently under extreme conditions significant quantities of H can be dissolved in molten Fe [25]. However, experimental data on steel reduced by \ch{H2} have proven that H escapes from molten Fe during solidification in a significantly fast manner [26]. The reason is attributed to relatively small value of H's migration enthalpy in solid steel which is 0.01 eV [27] (compared to that of Cr which is 0.761 eV or that of Ni which is 1.338 eV [28]). Consequently, the H content that is dissolved during interaction with polymers is expected to escape out of solid steel in the dry environment over macroscopic time scales after solidification. This point can be further verified by the fact that at room temperature, the diffusivity of H in pure bcc Fe is about $1.0 \times 10^{-4}$ cm$^2$ s$^{-1}$ compared to $1.0 \times 10^{-16}$ cm$^2$ s$^{-1}$ for C [7]. As a result, H's higher diffusivity implies that in the absence of trapping centers such as dislocations and point defects, H's concentration will decrease to very small quantities over matter of hours in homogeneous steel. However, depending on factor such as the cold-working and quenching process, H may be trapped in dislocations. For instance, the concentration of H is expected to be $10^{19}$ cm$^{-3}$ for dislocation density of $10^{11}$ cm$^{-2}$ [29].

       \section{CONCLUSIONS}\FloatBarrier
Using \textit{ab initio} molecular dynamics, we could quantitatively study the interaction between polyurethane and polysulfide monomers with molten Fe. The results can be summarized as following:\\
1.  The dissolved carbon mass ratio from polyurethane and polysulfide were $0.49 \%$ and $1.68 \%$ respectively. In this regard polysulfide is a superior carburizer than polyurethane. Nonetheless, both of these polymers can be used as carburizing agents instead of natural carbonaceous materials such as fossil fuels and coal.\\
2.  H diffusivity in molten Fe was detected. Such diffusivity has been observed in earlier experiments and it is known to be mostly reversible and critically temperature, pressure and process dependent.\\
3.  Within the range of the simulation time, no diffusion of S or N into the bulk region of molten Fe was observed. This point was in agreement with previous experiments reporting the accumulation of these elements at the interface. In the case of S, it is expected to be captured by the slag.\\
4.  Polyurethane interaction with molten Fe resulted in emitted \ch{H2} molecules while polysulfide interaction resulted in \ch{H2}, S and \ch{CH_x}. The emitted \ch{H2} and \ch{CH_x} can be utilized as heat resources or reducing agents depending on the application. \par
To conclude, these simulations, although computationally expensive, shed light on interesting phenomenon regarding the interaction of polymeric monomers with molten Fe. The anticipated advances in computer power combined with more efficient software parallelization in near future will mean longer and larger simulation will be feasible to perform in order to fully understand the mechanisms governing the dissolution of polymer in molten metals.
             
       \section{CONFLICTS OF INTEREST}
The authors declare that there is no conflict of interest.
       
      \section{ACKNOWLEDGMENTS}
This work was supported by Australian Research Council through Grant No. FT0992021. The computational facility was provided by Intersect Australia Ltd.

       \section{References}
[1] M. Meier, Sustainable polymers: reduced environmental impact, renewable raw materials and catalysis, Green Chem. 16 (2014) 1672--1672. \par
[2] J. Hopewell, R. Dvorak and E. Kosior, Plastics recycling$\colon$ challenges and opportunities, Philos. T. R. Soc. B 364 (2009) 2115--2126. \par
[3] D.K.A. Barnes, F. Galgani, R.C. Thompson and M. Barlaz, Accumulation and fragmentation of plastic debris in global environments, Philos. T. R. Soc. B 364 (2009) 1985--1998. \par
[4] National Waste Policy: Less waste, More Resources Implementation Report Australian Government, Department of the Enviornment, (2012) \url{http://www.scew.gov.au/system/files/resources/7b8074ab-ea58-4981-9c89-ccdfa49246ad/files/national-waste-policy-implementation-report-2011.pdf}. \par
[5] Container and Packaging Recycling Law, Ministry of the Environment, Japan, (2005) \url{http://www.env.go.jp/en/laws/recycle/07.pdf}. \par
[6] Public consultation on the Green Paper on Plastic Waste, European Commission, Directorate General for the Environment, (2013) \url{http://ec.europa.eu/environment/consultations/plastic_waste_en.htm}. \par
[7] M. Asanuma, T. Ariyama, M. Sato, R. Murai, T. Nonaka, I. Okochi, H. Tsukiji and K. Nemoto, Development of waste plastics injection process in blast furnace, ISIJ Int. 40 (2000) 244--251. \par
[8] V. Sahajwalla, M. Zaharia, S. Kongkarat, R. Khanna, M. Rahman, N. Saha-Chaudhury, P. O'Kane, J. Dicker, C. Skidmore and D. Knights, Recycling End-of-Life Polymers in an Electric Arc Furnace Steelmaking Process: Fundamentals of Polymer Reactions with Slag and Metal, Energy Fuels 26 (2012) 58--66. \par
[9] K.M. Zia, H.N. Bhatti and I. Ahmad Bhatti, Methods for polyurethane and polyurethane composites, recycling and recovery: A review, React. Funct. Polym. 67 (2007) 675--692. \par
[10] Dirk Lechtenberg, Tyres as an alternative fuel, (2011) \url{http://www.cemfuels.com/articles/318-tyres-as-an-alternative-fuel}, Accessed on 1st of May 2014. \par
[11] G. Kresse and J. Hafner, Ab-initio Molecular-dynamics Simulation of the Liquid-metal Amorphous-semiconductor Transition in Germanium, Phys. Rev. B 49 (1994) 14251--14269. \par
[12]	G. Kresse and J. Furthmuller, Efficient iterative schemes for \textit{ab initio} total-energy calculations using a plane-wave basis set, Phys. Rev. B 54 (1996) 11169--11186. \par
[13] G. Kresse and D. Joubert, From ultrasoft pseudopotentials to the projector augmented-wave method, Phys. Rev. B 59 (1999) 1758--1775. \par
[14] J.P. Perdew, J.A. Chevary, S.H. Vosko, K.A. Jackson, M.R. Pederson, D.J. Singh and C. Fiolhais, Atoms, Molecules, Solids, and Surfaces - Applications of the Generalized Gradient Approximation for Exchange and Correlation, Phys. Rev. B 46 (1992) 6671--6687. \par
[15] S. Nose, A Unified Formulation of The Constant Temperature Molecular-Dynamics Methods, J. Chem. Phys. 81 (1984) 511--519. \par
[16] W.G. Hoover, Canonical Dynamics - Equilibrium Phase-space Distributions, Phys. Rev. A 31 (1985) 1695--1697. \par
[17] M.H.N. Assadi and V. Sahajwalla, Recycling End-of-Life Polycarbonate in Steelmaking: Ab Initio Study of Carbon Dissolution in Molten Iron, Ind. Eng. Chem. Res. 53 (2014) 3861--3864. \par
[18] J.D. Katz and T.B. King, The Kinetics of Nitrogen Absorption and Desorption from A Plasma-Arc by Molten Iron, Metall. Trans. B 20 (1989) 175--185. \par
[19] P. Mullner, C. Solenthaler, P. Uggowitzer and M.O. Speidel, On The Effect of Nitrogen on the Dislocation-structure of Austenitic Stainless-steel, Mat. Sci. Eng. A-Struct. 164 (1993) 164--169. \par
[20] R. Khanna, R. Mahjoub and V. Sahajwalla, in Applications of Monte Carlo Method in Science and Engineering, edited by S. Mordechai (InTech Europe, Rijeka, Croatia, 2011), pp. 641--645. \par
[21] R.G. Olsson, V. Koump and T.F. Perzak, Rate of Dissolution of Carbon in Molten Fe-C Alloys, T. Metall. Soc. AIME 236 (1966) 426--429. \par
[22] M.Y. Solar and R.I.L. Guthrie, Hydrogen Transport In Stagnant Molten Iron, Metall. Trans. 2 (1971) 457--464. \par
[23] G.-R. Jiang, Y.-X. Li and Y. Liu, Calculation of hydrogen solubility in molten alloys, T. Nonferr. Metal. Soc. China 21 (2011) 1130--1135. \par
[24]	M.Y. Solar and R.I.L. Guthrie, Hydrogen Diffusivities in Molten Iron Between 0.1 And 1.5 Atm \ch{H2}, Metall. Trans. 2 (1971) 3241--3242. \par
[25] T. Okuchi, Hydrogen partitioning into molten iron at high pressure: Implications for Earth's core, Science 278 (1997) 1781--1784. \par
[26] Y. Kashiwaya and M. Hasegawa, Thermodynamics of Impurities in Pure Iron Obtained by Hydrogen Reduction, ISIJ Int. 52 (2012) 1513--1522. \par
[27] D.E. Jiang and E.A. Carter, Diffusion of interstitial hydrogen into and through bcc Fe from first principles, Phys. Rev. B 70 (2004) 064102. \par
[28] T. Tsuru and Y. Kaji, First-principles thermodynamic calculations of diffusion characteristics of impurities in gamma-iron, J. Nucl. Mater. 442 (2013) S684--S687. \par
[29] R.A. Oriani, Diffusion and Trapping of Hydrogen in Steel, Acta Metall. 18 (1970) 147--157.

\end{document}